\documentclass[journal]{IEEEtran}
\usepackage{lineno}
\usepackage{graphicx}
\usepackage{soul,color}
\usepackage{adjustbox}
\usepackage{multirow}
\usepackage{amsmath} 
\usepackage{microtype}
\usepackage{url}
\usepackage{subcaption}

\begin{document}
\title{Exploring Challenges in Deep Learning of Single-Station Ground Motion Records}

\author{
    \IEEEauthorblockN{
        Ümit Mert Çağlar\IEEEauthorrefmark{1}, 
        Baris~Yilmaz\IEEEauthorrefmark{1}, 
        Melek~Türkmen\IEEEauthorrefmark{1}, Erdem~Akagündüz\IEEEauthorrefmark{1},
        Salih~Tileylioglu\IEEEauthorrefmark{2}
    } \\
    \IEEEauthorblockA{\IEEEauthorrefmark{1}\textit{Dept. of Modeling and Simulation, Graduate School of Informatics,\\ Middle East Technical University, Ankara, Türkiye}}\\
    \IEEEauthorblockA{\IEEEauthorrefmark{2}\textit{Dept. of Civil Engineering, Kadir Has University, Istanbul, Türkiye}}

    \IEEEauthorblockN{
        \{mecaglar,
yilmaz.baris\_01,
turkmen.melek,
akaerdem\}@metu.edu.tr, salih.tileylioglu@khas.edu.tr
    }
}

\maketitle
\begin{abstract}
Contemporary deep learning models have demonstrated promising results across various applications within seismology and earthquake engineering. These models rely primarily on utilizing ground motion records for tasks such as earthquake event classification, localization, earthquake early warning systems, and structural health monitoring. 
However, the extent to which these models truly extract meaningful patterns from these complex time-series signals remains underexplored.
In this study, our objective is to evaluate the degree to which auxiliary information, such as seismic phase arrival times or seismic station distribution within a network, dominates the process of deep learning from ground motion records, potentially hindering its effectiveness.
Our experimental results reveal a strong dependence on the highly correlated Primary (P) and Secondary (S) phase arrival times. These findings expose a critical gap in the current research landscape, highlighting the lack of robust methodologies for deep learning from single-station ground motion recordings that do not rely on auxiliary inputs.
\end{abstract}


\section{Introduction}
At the core of deep learning lies a defining element: deep features. These abstract representations, automatically extracted through multiple layers of a neural network, enable the model to discern intricate patterns and properties within a complex input signal. The depth of the network allows it to learn increasingly complex and nuanced representations, empowering it to make more informed and accurate predictions. In essence, deep features serve as the bedrock of the remarkable capabilities exhibited by deep learning models in tasks such as image recognition \cite{he2016residual}, natural language processing \cite{clip}, and various other domains, including seismology and earthquake engineering \cite{Mousavi2020b}.
 
Earthquakes occur when faults rupture underground. The location where the rupture occurs is termed as the focus of the earthquake. Once a fault ruptures, waves propagate through the rocks, traversing soils along their trajectory, until they eventually reach the ground surface. The first waves to reach the ground surface are the body waves (P and S waves, respectively). These waves are then followed by the surface waves, known as Rayleigh and Love waves. The waves that reach the surface create the ground motions felt during earthquakes. To study earthquakes and their impacts, seismic stations are established with the purpose of recording ground motions resulting from seismic activity. Ground motions caused by an earthquake are generally defined by three components perpendicular to each other. Seismic stations generally include a sensor usually capable of sensing the signal in three directions as well as a recorder. Depending on the size of the earthquake and proximity to the fault, ground motion amplitudes caused by an earthquake can range from a few nanometers to several meters, and thus, it becomes a challenge to record the wide range of signals with a single type of instrument \cite{Alguacil2015}. Earthquake measuring instruments come in various types, each tailored for specific purposes, characterized by distinct dynamic ranges and bandwidths. Stations consisting of seismographs, which serve as the primary data sources for this study, play an important role in determining an earthquake's location. 

The location of an earthquake is commonly defined by its epicenter, which is the projection of the focus of the earthquake on the ground surface. The epicentral distance, the focal point of this study, refers to the distance between the epicenter of an earthquake and a specific location, typically a seismic station. This distance is typically estimated by measuring the time difference between the arrival of P waves and S waves at a station. 

The convergence of deep learning with seismology and earthquake engineering is a recent development in the literature. Based on a rather small-scale dataset consisting of roughly $\sim$3k events, \cite{Perol2018} established the first DL-based earthquake engineering model in 2018. In the short period that followed this seminal study, there have been various studies in seismology and earthquake engineering that used deep learning for different purposes, such as in 
event detection for early warning \cite{Kuyuk2018,Lomax2019,Mousavi2020b,Münchmeyer2020,Yano2021,Bilal2021},
event classification \cite{Kim2022,Nakano2022}, 
ground response estimation \cite{Hong2021},
earthquake (EQ) phase picking \cite{Pardo2019,Mousavi2020b},
magnitude estimation \cite{Jozinović2020,Ende2020,Münchmeyer2020,Zhang2021,Ristea2022,Saad2021,Bloemheuvel2022} \cite{sadhukhan2023predicting}, 
EQ origin time estimation \cite{Mousavi2020a,Saad2021}, 
epicenter location classification \cite{Kuyuk2018,Lomax2019,Kriegerowski2019,Saad2021}, epicentral distance estimation \cite{Ristea2022,Yoma2022}, soil parameter estimation \cite{Yilmaz2024},
depth prediction \cite{Kriegerowski2019,Mousavi2020a,Ende2020,Zhang2021,Ristea2022,Saad2021,Bilal2021}, 
and epicenter coordinates prediction \cite{Zhang2020,Mousavi2020a,Ende2020,Zhang2021,Bilal2021}. In recent studies, deep learning methods have been used for scalable and explainable earthquake magnitude prediction\cite{singh2025explainable}, forecasting earthquake precursor signals by monitoring ionospheric anomalies\cite{budak2023lstm}, which are often associated with earthquakes \cite{eroglu2023ionospheric}. Detecting and evaluating P waves is a crucial step for early warning systems \cite{chandrakumar2025evaluating} while estimating the S wave by using the P wave signals contribute further to these systems \cite{chandrakumar2024estimating}. There are also tools to improve the highly noisy nature of the seismological signals, such as the Convex Non-Convex Fused Lasso Signal Approximation (CNC-FLSA)\cite{goudarzi2025improving} and improved wavelet mode maximum value (IMWMM)\cite{yao2023novel}.

A large-scale dataset is essential to any deep learning study, especially when it comes to creating a high-level (i.e., deep) representation of the input signal. However, in the field of AI-based seismology and earthquake engineering, despite its importance, such a corpus containing hundreds of thousands of EQ records did not exist until very recently. This necessity led to the publication of the Stanford Earthquake Dataset (STEAD) by \cite{Mousavi2019}, which contained more than 1M seismic waveforms. The same group used a subset of this dataset in their subsequent studies, first using a Bayesian DNN to detect P-arrival times and localize epicenters \cite{Mousavi2020a}, and then using a transformer and LSTM-included architecture for event and phase picking \cite{Mousavi2020b}. STEAD was further investigated by another group \cite{Ristea2022} that employed a complex CNN to address earthquake localization and magnitude estimation problems.
The significance of \cite{Ristea2022} lies in their distinct approach to predicting epicentral distances without relying on auxiliary data such as P/S phase arrival information, in contrast to \cite{Mousavi2020a} which utilizes this auxiliary data in their method. However, \cite{Ristea2022}'s pretrained model remains undisclosed to the public. Although their code repository is accessible, the authors have provided limited information about the subset of the STEAD set used in their experiments, which poses challenges for reproducing their results.

\begin{figure}[t]
    \centering
    \includegraphics[width=1\linewidth]{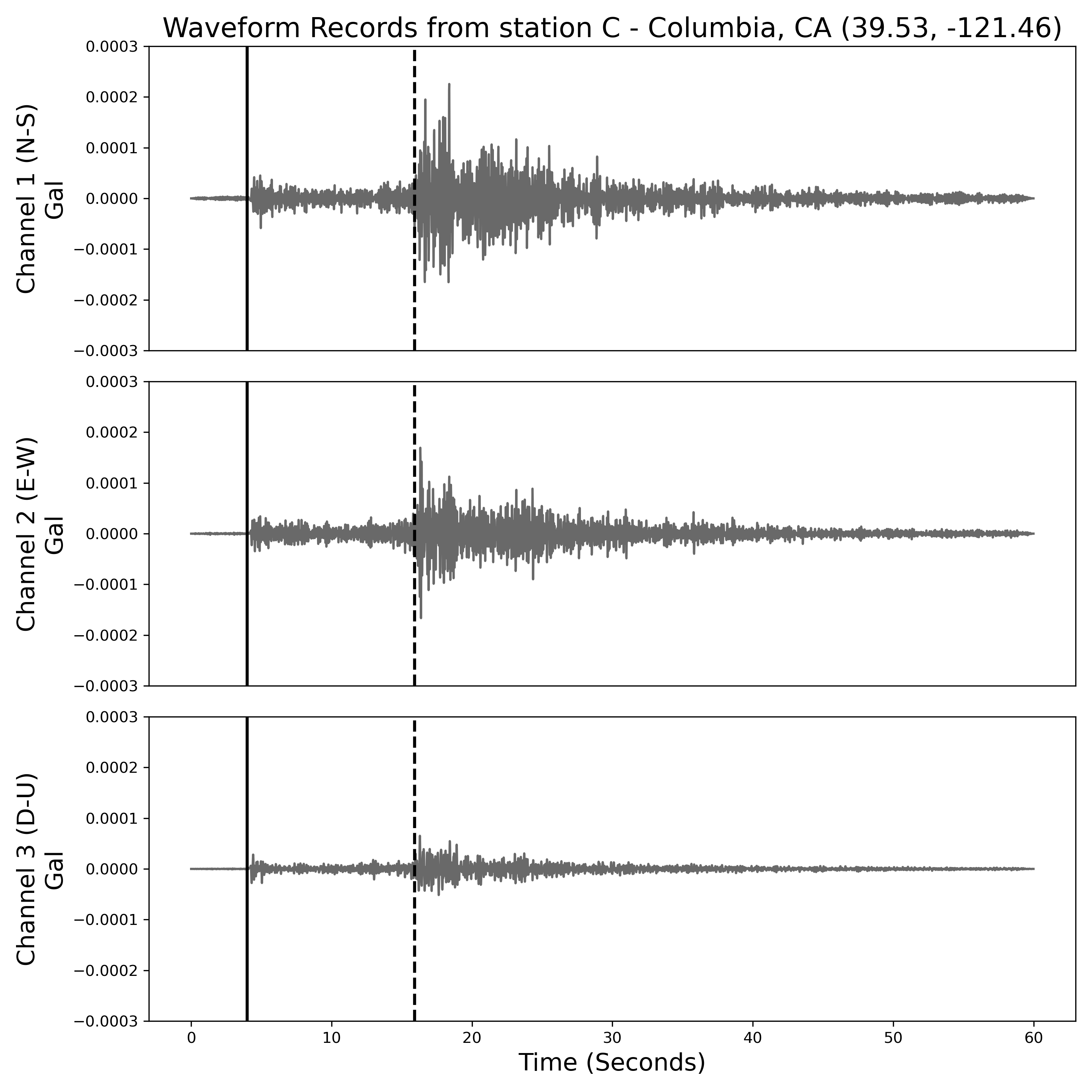}
    \caption{A sample recorded event from the STEAD, belonging to station at 38.034, -120.38, Columbia (California). Primary (solid) and Secondary wave (dashed) arrival time indices are depicted as dashed lines.}
    \label{fig:stead_sample}
\end{figure}

The historical evolution of deep learning unfolds a recurring theme that starts with crafting an encoder \cite{sarker2021deep}. From raw signals, this encoder can extract deep and nontransparent representations. The recent developments in the field show that the encoder then becomes an off-the-shelf tool, ready for deployment in subsequent downstream tasks once this milestone is reached, such as ResNet \cite{he2016residual} for vision. As far as seismology and earthquake engineering are concerned, an important question arises:
``\emph{are there any off-the-shelf encoders that can extract transferable seismic deep representations}''?

To address this question, we take a step back to re-implement the most promising deep learning architectures from the literature, namely the Residual Networks (ResNet) \cite{he2016residual} and Temporal Convolutional Networks (TCN) \cite{TCN2016} and conduct experiments on STEAD to catch the state-of-the-art results in this domain. We then apply an ablation study to find the effects of different hyperparameters and most importantly, the effects of auxiliary information, such as the primary/secondary (P/S) phase signal arrival times (P/S phase information from hereafter). \cite{Kuyuk2018}.
Previous studies that attempted to localize epicenters from single station ground motion data \cite{Mousavi2020a,Mousavi2024} used P/S phase information with ground motion records. Alternatively, some deep-learning studies use multiple station information 
\cite{Kriegerowski2019} \cite{Zhang2020} \cite{Ende2020} \cite{Zhang2021} \cite{Bilal2021} when processing the ground motion records for different down-stream tasks. In this paper, we conduct a set of experiments to observe ``how deeply'' we could learn from only ground motion records without using auxiliary information such as P/S phase arrival times.

We hypothesize that deep learning models may overfit to highly correlated auxiliary information, such as P/S phase arrival times, potentially limiting their ability to learn from the waveform data itself when predicting epicentral distance. To investigate this, we focus specifically on epicenter localization using accelerometer records from a \emph{single} station, thereby minimizing the influence of auxiliary inputs like station distribution or phase arrival. Using the STEAD dataset, we conduct a comprehensive hyperparameter search to identify the most effective model configurations. To ensure transparency and reproducibility, we provide not only our code repository but also the full history of our experiments through a publicly accessible project framework.

\begin{figure}[t]
    \centering
    \includegraphics[trim={15 12 0 0},clip,width=0.6\linewidth]{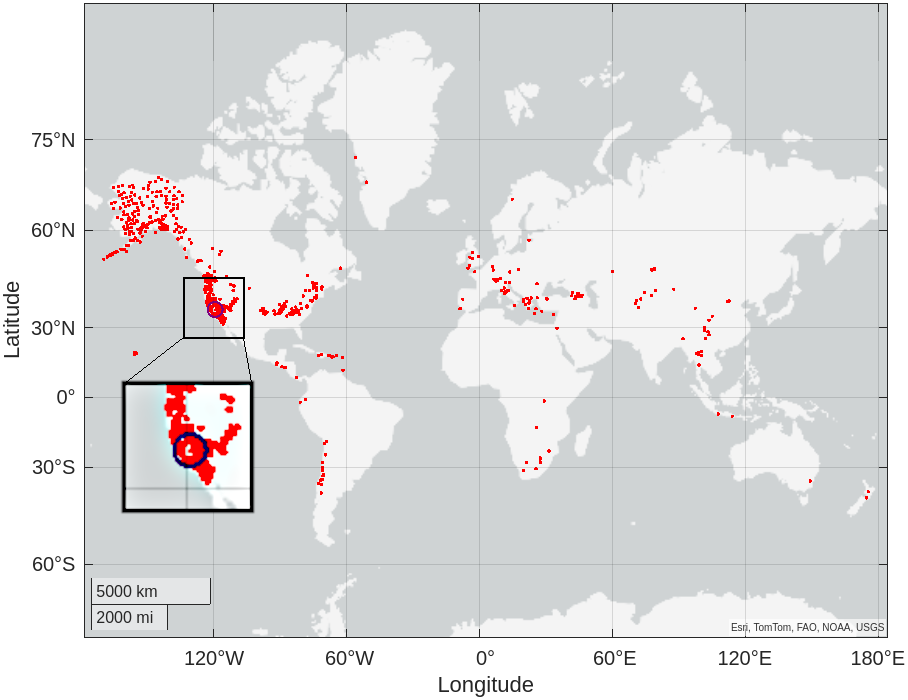}
    \caption{Station locations extracted from the benchmarked subset of the STEAD utilized in \cite{Mousavi2020a}. The circled area indicates the local subset focused on a 300km radius centered around the California state.}
    \label{fig:stations}
\end{figure}

\begin{figure*}[t]
    \centering
    \includegraphics[trim={0 0 0 0},clip,width=0.65\linewidth]{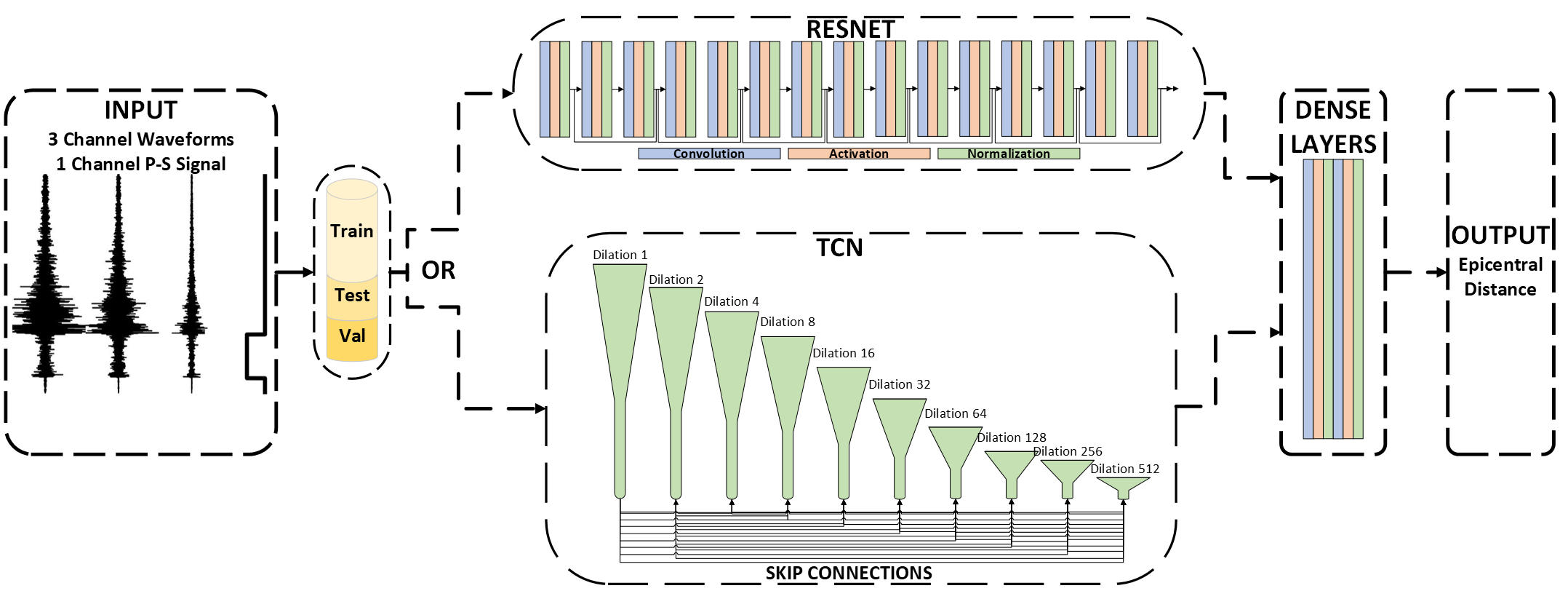}
    \caption{Overall system architecture with optional P-S channel in addition the 3-channel earthquake waveform, variable size dense layers, ResNet or TCN for the encoder.}
    \label{fig:models}
\end{figure*}

The remainder of this paper is organized as follows. Methodology section includes details of the STEAD dataset and the deep learning models employed. Experiments section describes the setup, the ablation study design, and our experiment tracking framework, which includes a public dashboard and summary reports. Then we present the best-performing hyperparameter and model configurations, followed by a comprehensive analysis of the results. This includes a discussion on the correlation between P/S phase arrival times and epicentral distances, as well as visualizations of model predictions and training curves. Finally, we conclude the paper with a summary of key findings and their implications for future research.

\section{Methodology}

\subsection{Dataset}

STanford EArthquake Dataset (STEAD): A Global Data Set of Seismic Signals for AI \cite{Mousavi2019} contains 60 seconds long seismic activities associated with roughly 450,000 earthquakes. The dataset contains roughly 1 million events between January 1984 and August 2018. The records have 6000 samples for each event corresponding to 100 Hz recording frequency. STEAD dataset is the main dataset that forms the baseline for this work. A sample that represents an event of an earthquake is provided in Figure \ref{fig:stead_sample}. It consists of ground motion records in North-South, East-West, and Up-Down dimensions, as well as labels for the Primary (P) and Secondary (S) signal arrival time.

In the original dataset, there are 2613 stations; however, in the studies associated with the dataset \cite{Mousavi2020a}, authors filter out waveforms that are generated by earthquakes further than 110 km or signal-to-noise (SNR) ratio of 25dB and lower. They have also omitted the stations where north-south and east-west components are not aligned to their correct geographic orientations. With these filters, 147,195  records can be used for epicenter distance estimation from single-station records. 
We kept the same filtering approach as in \cite{Mousavi2020a}
and utilized the aforementioned subset with 743 stations that record earthquakes all over the globe, as depicted in Figure \ref{fig:stations}. 
Utilizing this STEAD subset as our foundation, which we refer to as the ``global'' subset in this paper, we further refine our analysis by establishing a more localized subset focused on a 300km radius centered around California, which we term the ``local'' subset (depicted as a circle in Figure \ref{fig:stations}). We conduct distinct experiments on these subsets to discern differences in the depth of features extracted from each. We chose to utilize the identical training set (80\%) and test set (20\%) as described in \cite{Mousavi2020a}. The rationale behind this decision lies in the fact that \cite{Mousavi2020a} represents one of the few large-scale deep learning experiments that rely solely on single station records without incorporating network topology information. By adopting the same subset of STEAD, we aim to conduct an ablation study focusing on a controlled auxiliary factor, specifically the phase information.

\subsection{Deep Learning Models}

In this study, we benchmark two fundamental CNN architectures. The first one is the ResNet \cite{he2016residual}, which is the most widely used CNN encoder architecture that proved to be effective on various data and several down-stream tasks in the literature. ResNet architecture enables deeper networks to be trained for more complex signals that require deeper understanding via residual connections. These residual connections enable the flow of gradients throughout the network, effectively eliminating the problematic deep learning phenomena of vanishing gradients, which disables deeper networks from being trained. We also experiment on the Temporal Convolutional Network (TCN) \cite{TCN2016}, which employs a convolutional model in a sequential (time-series) manner. Convolutional networks are known for their efficient performance on data with spatial information, such as images where pixels in close proximity are highly correlated. Similarly, in time-series signals, such as earthquake waveforms, the value of the waveform at time "t" is correlated with the neighboring time steps. Furthermore, TCN has been used in experiments in \cite{Mousavi2020a}, which address the same problem definition as ours.
The general model architectures are depicted in Figure \ref{fig:models}.

\section{Experiments}

We utilized exhaustive hyperparameter search experiments to identify the best-performing deep learning scenario, both with and without the inclusion the P/S phase information. For this purpose, we have used a grid search method with two different datasets, three different learning rates, two different gamma values, and three different architectural parameters on two different deep learning models, resulting in at least 144 distinct experiments. For distance prediction, mean absolute error (L1 loss) is chosen and directly calculated using the haversine distance in kilometers between the epicenter and the recording station. 

\subsection{Ablation}

The ablation study of this work aims to evaluate the importance of the P/S phase arrival time information, which is highly correlated with the distance of the epicenter from the station. This investigation aims to determine whether deep learning models can learn from only the three-channel sensor signals or rely heavily on the P-S arrival time difference to extract the distance.
We posit that rather than learning from only ground motion records, the models may predominantly rely on the highly correlated P-S arrival times for the extraction of relative epicentral distances.
To achieve this objective, our ablation study incorporates a fundamental distinction: the inclusion of the P/S phase information. We created and trained two identical setups, differing solely in the presence or absence of the P/S signal information.

\subsection{Experiment Tracking}

The experiments conducted in this work were monitored through Weights and Biases (WANDB) framework. The experiments can be followed through a publicly available and online \cite{caglar2024wandb}. The report includes details such as the total runtime of experiments, parameter importance concerning the network model (TCN-ResNet), learning rate, sizes of fully connected layers, gamma values, and most notably, the presence of the P/S phase information as input.

\subsection{Resources and Experimental Costs}

The reproducible experiments in this work takes 360 Hours to complete on an A100 GPU with 80 GB of VRAM, 16 cores of AMD 7742 server CPUs with 200 GB RAM. With a rated peak power of A100 of 400Watts and 225 Watts for the average CPU, our experiments had one-fourth CPU cores resulting in 55 Watts of average CPU usage and mean GPU power usage statistics recorded as 225 Watts, resulting in 280 Watts for only GPU and CPU power consumption. Other components such as RAM and SSD or hard drives would cost roughly 20 Watts for our experiments, resulting in a total of 300 Watts. With the total amount of training of 268 hours for TCN experiments and 92 hours for ResNet, the grand total power consumption of our work was 108 KWh. We believe that our experiments, models, and approach are sustainable and have minimal impact on the environment due to the compact nature of our deep learning models and optimized training structure. The best-performing model (TCN) we have trained throughout this work takes about 300KB of memory and trains in 100 minutes for the local subset and 400 minutes for the global dataset.

\section{Results}

To structure our findings effectively, we have categorized our experimental results into two primary subgroups: ResNet and TCN experiments. In the following, we first provide a comprehensive summary of each experiment and proceed to showcase our prediction charts and learning plots from the experiments conducted both with and without the inclusion of the P/S phase information as input. Moreover, we explore the correlation between P/S phase arrival differences and epicentral distances to enrich our analyses.

\begin{figure}[t]
    \centering
    \includegraphics[width=1\linewidth,trim={60pt 0 70pt 0},clip]{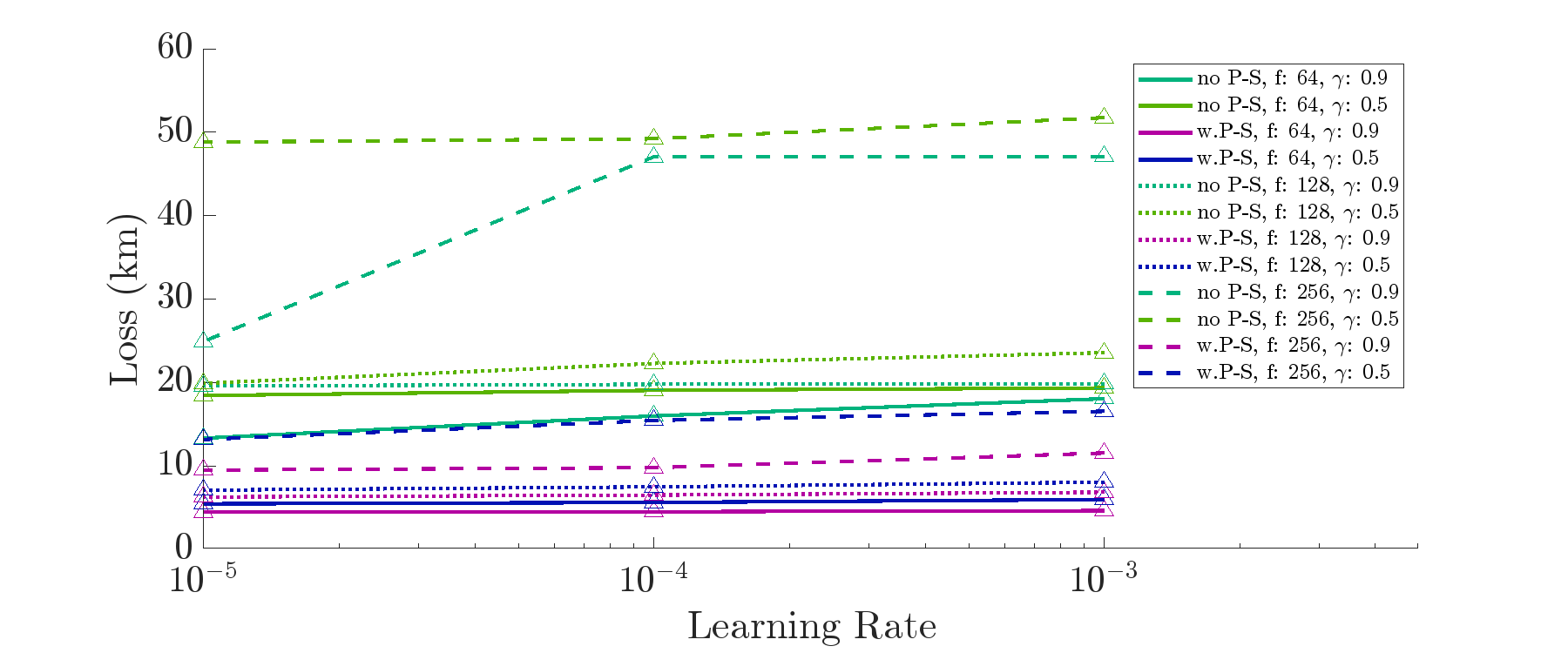}
    \caption{Test set average loss values for different hyperparameters, with and without the inclusion of the P/S phase information, on the \textbf{ResNet} model using the \textbf{local} subset.}
    \label{ResNet_Local}
\end{figure}
\begin{figure}[t]
    \centering
    \includegraphics[width=1\linewidth,trim={60pt 0 70pt 0},clip]{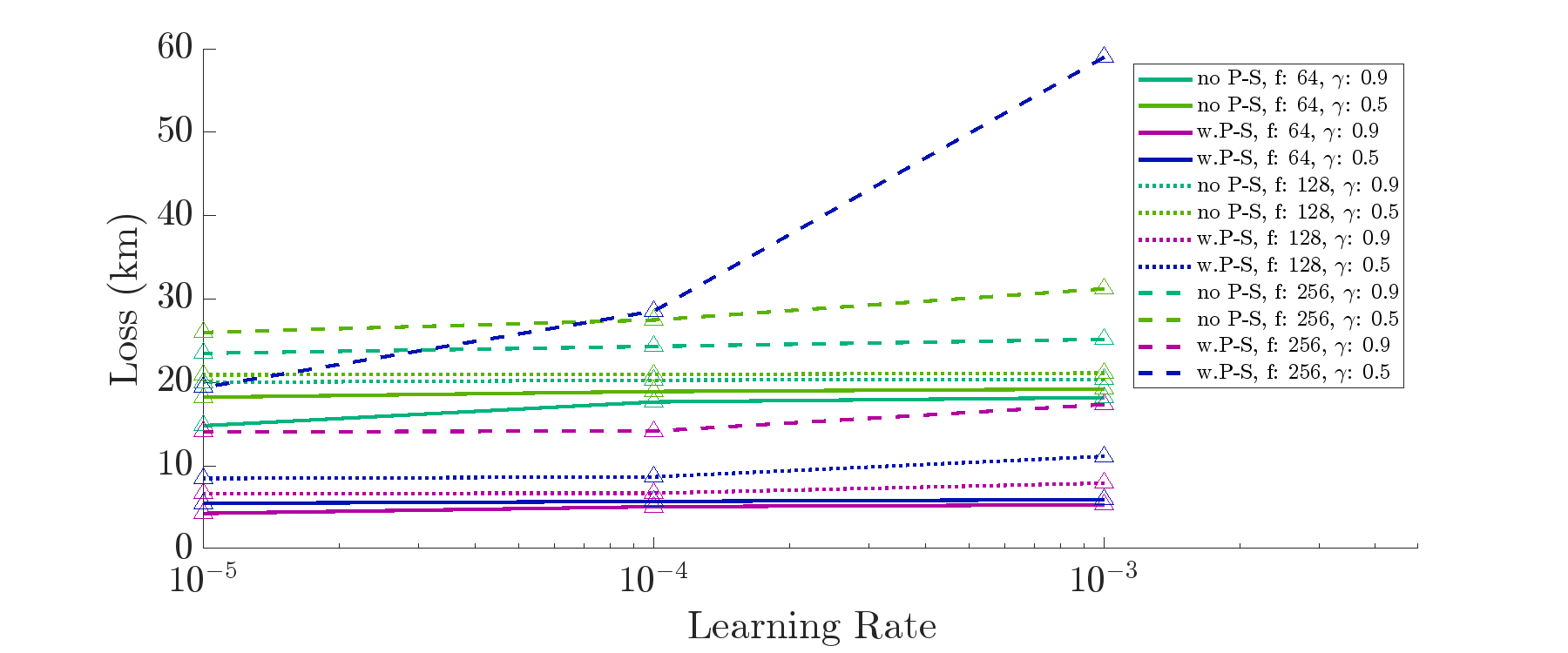}
    \caption{Test set average loss values for different hyperparameters, with and without the inclusion of the P/S phase information, on the \textbf{ResNet} model using the \textbf{global} subset.}
    \label{ResNet_Global}
\end{figure}
\begin{figure}[ht!]
    \centering
    \includegraphics[width=1\linewidth,trim={60pt 0 70pt 0},clip]{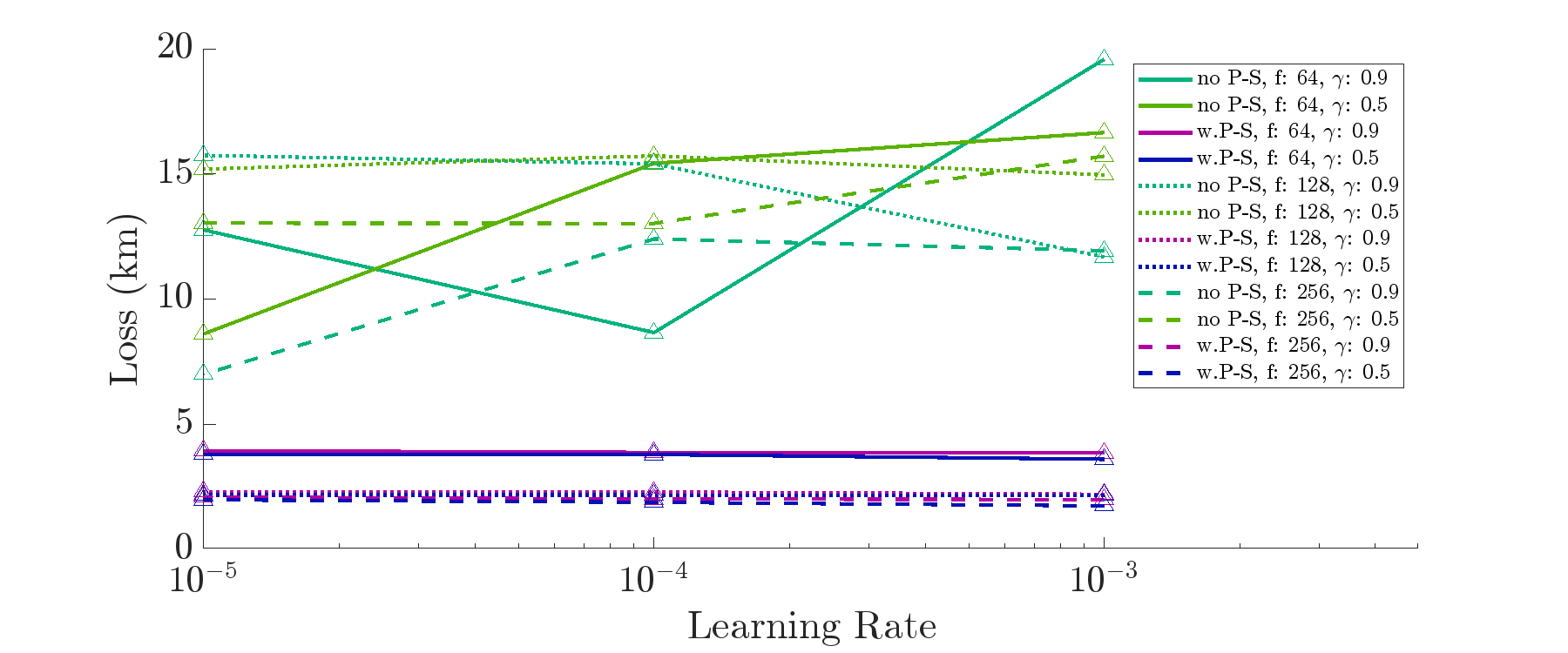}
    \caption{Test set average loss values for different hyperparameters, with and without the inclusion of the P/S phase information, on the \textbf{TCN} model using the \textbf{local} subset.}
    \label{TCN_local}
\end{figure}
\begin{figure}[ht!]
    \centering
    \includegraphics[width=1\linewidth,trim={60pt 0 70pt 0},clip]{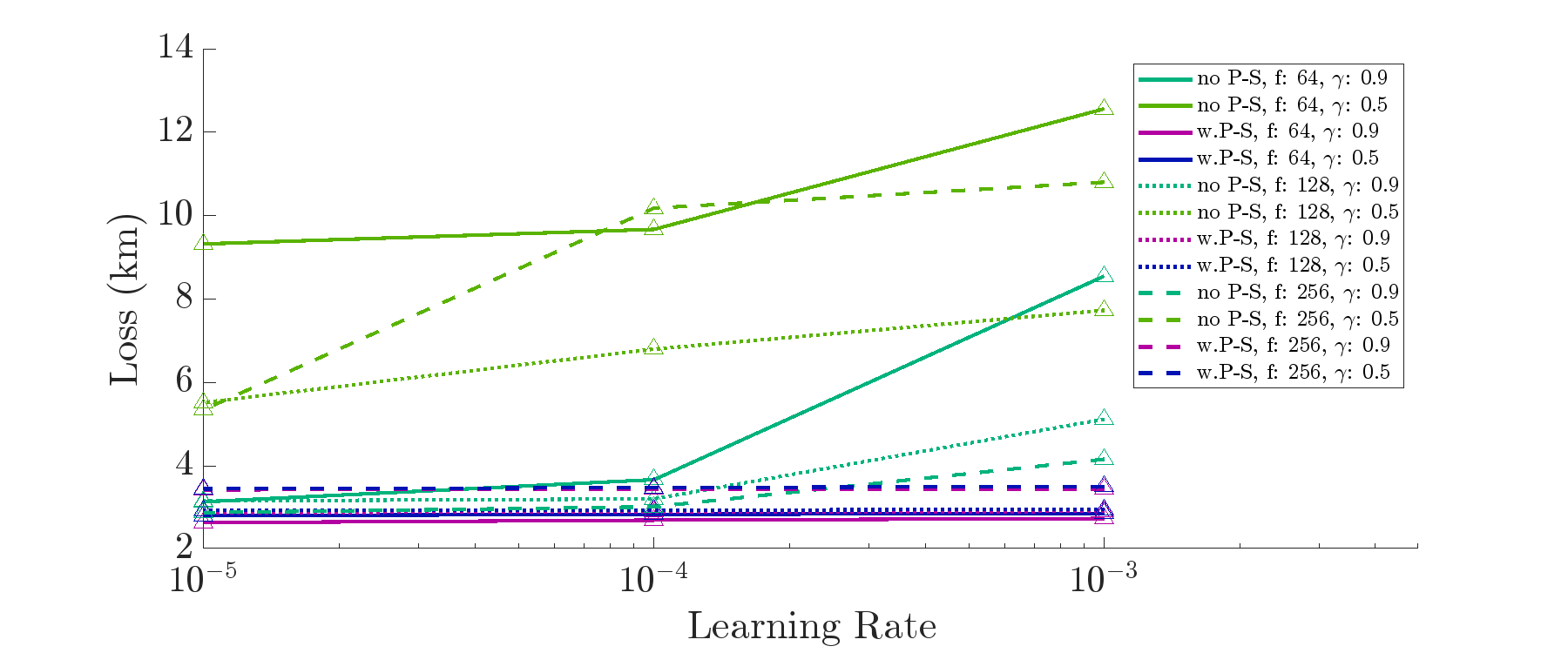}
    \caption{Test set average loss values for different hyperparameters, with and without the inclusion of the P/S phase information, on the \textbf{TCN} model using the \textbf{global} subset.}
    \label{TCN_Global}
\end{figure}

\subsection{Experiments with the ResNet Architecture}

Experimental results for the ResNet model trained with the local subset centered around California, with and without using P/S phase information, are given in Table \ref{experiments_resnet_1}. The best-performing hyperparameter combination is shown in bold fonts and has an L1 Loss score of 13.33 km without phase information and 4.47 km with it. The same results are depicted in Figure \ref{ResNet_Local}. 

\begin{table}[ht!]
    \centering
    \begin{tabular}{|c|c|c|c|c|}
        \hline
        \multicolumn{5}{|c|}{ResNet Local} \\
        \hline
        Size & $\gamma$ & lr & No PS & PS\\
        \hline
        256 & 0.5 & 0.001     & 51.74  & 16.55 \\ \hline
        256 & 0.5 & 0.0001    & 49.21  & 15.45 \\ \hline
        256 & 0.5 & 0.00001   & 48.81  & 13.17 \\ \hline
        256 & 0.9 & 0.001     & 47.08  & 11.55 \\ \hline
        256 & 0.9 & 0.0001    & 47.06  & 9.79  \\ \hline
        256 & 0.9 & 0.00001   & 24.93  & 9.51  \\ \hline
        128 & 0.5 & 0.001     & 23.57  & 8.05  \\ \hline
        128 & 0.5 & 0.0001    & 22.28  & 7.49  \\ \hline
        128 & 0.5 & 0.00001   & 19.9   & 7.1   \\ \hline
        128 & 0.9 & 0.001     & 19.86  & 6.85  \\ \hline
        128 & 0.9 & 0.0001    & 19.78  & 6.51  \\ \hline
        128 & 0.9 & 0.00001   & 19.58  & 6.28  \\ \hline
        64  & 0.5 & 0.001     & 19.35  & 5.97  \\ \hline
        64  & 0.5 & 0.0001    & 19.06  & 5.62  \\ \hline
        64  & 0.5 & 0.00001   & 18.44  & 5.46  \\ \hline
        64  & 0.9 & 0.001     & 18.06  & 4.64  \\ \hline
        64  & 0.9 & 0.0001    & 15.99  & 4.51  \\ \hline
        64  & 0.9 & 0.00001   & 13.33  & \textbf{4.47}  \\ \hline
Best:& 4.47 & $\mu\pm\sigma:$ & $27.67\pm13.74$ & $8.28\pm3.69$  \\ \hline
    \end{tabular}
    \caption{Experimental results for the Resnet model trained with and without P/S phase information on the local subset}
    \label{experiments_resnet_1}
\end{table}

The results for the ResNet model trained with, this time, the global dataset with and without the inclusion of the P/S phase information as input are given in Table \ref{experiments_resnet_2}. Again, the best-performing hyperparameter combination is shown in bold font and has L1 Loss score of 14.82 km without the phase information and 4.31 km with it. The results of the experiments are also depicted in Figure \ref{ResNet_Global}.

\begin{table}[ht!]
    \centering
    \begin{tabular}{|c|c|c|c|c|}
        \hline
        \multicolumn{5}{|c|}{ResNet Global} \\
        \hline
        Size & $\gamma$ & lr & No PS & PS\\
        \hline
        256 & 0.5 & 0.001     & 31.23  & 58.99 \\ \hline
        256 & 0.5 & 0.0001    & 27.49  & 28.56 \\ \hline
        256 & 0.5 & 0.00001   & 26     & 19.41 \\ \hline
        256 & 0.9 & 0.001     & 25.17  & 17.36 \\ \hline
        256 & 0.9 & 0.0001    & 24.32  & 14.18 \\ \hline
        256 & 0.9 & 0.00001   & 23.49  & 14.09 \\ \hline
        128 & 0.5 & 0.001     & 21.13  & 11.15 \\ \hline
        128 & 0.5 & 0.0001    & 20.96  & 8.69  \\ \hline
        128 & 0.5 & 0.00001   & 20.92  & 8.49  \\ \hline
        128 & 0.9 & 0.001     & 20.38  & 7.94  \\ \hline
        128 & 0.9 & 0.0001    & 20.28  & 6.73  \\ \hline
        128 & 0.9 & 0.00001   & 20.01  & 6.7   \\ \hline
        64  & 0.5 & 0.001     & 19.21  & 5.96  \\ \hline
        64  & 0.5 & 0.0001    & 18.92  & 5.75  \\ \hline
        64  & 0.5 & 0.00001   & 18.23  & 5.51  \\ \hline
        64  & 0.9 & 0.001     & 18.19  & 5.34  \\ \hline
        64  & 0.9 & 0.0001    & 17.67  & 5.11  \\ \hline
        64  & 0.9 & 0.00001   & 14.82  & \textbf{4.31}  \\ \hline
Best:& 4.31 & $\mu\pm\sigma:$ & $21.58\pm4.02$  & $13.02\pm13.11$  \\ \hline	
    \end{tabular}
    \caption{Experimental results for the Resnet model trained with and without P/S phase information on the global dataset}
    \label{experiments_resnet_2}
\end{table}

The experiments in this section agree with our hypothesis that the inclusion of the P-S phase arrival time information as an additional input channel shows a direct effect on the capability of the model on predicting the epicentral distance from single station recordings. With the inclusion of the auxiliary P/S phase information, we obtain the best model L1 Loss scores of 4.47 km and 4.31 km, for local and global sets, respectively. The same respective results obtained without using P/S phase information are 13.33 km and 14.82 km, which are significantly higher in all iterations. 

\subsection{Experiments with the TCN Architecture}

The TCN model trained with the local dataset centered around California, with and without using P-S phase information, are provided in Table \ref{experiments1}. Again, the best-performing hyperparameter combination is shown in bold fonts and has L1 Loss score of 7 km for without the P/S phase information and 1.74 km with its inclusion as input. The same results are depicted in Figure \ref{TCN_local}. 

\begin{table}[ht!]
\centering
\begin{tabular}{|c|c|c|c|c|}
\hline
\multicolumn{5}{|c|}{TCN Local} \\
\hline
Size & $\gamma$ & lr & No PS & PS\\
\hline
256 & 0.5 & 0.001     & 15.71  & \textbf{1.74}  \\ \hline
256 & 0.5 & 0.0001    & 13.02  & 1.88  \\ \hline
256 & 0.5 & 0.00001   & 13.03  & 1.98  \\ \hline
256 & 0.9 & 0.001     & 11.93  & 1.99  \\ \hline
256 & 0.9 & 0.0001    & 12.4   & 2.01  \\ \hline
256 & 0.9 & 0.00001   & 7      & 2.09  \\ \hline
128 & 0.5 & 0.001     & 14.96  & 2.16  \\ \hline
128 & 0.5 & 0.0001    & 15.72  & 2.17  \\ \hline
128 & 0.5 & 0.00001   & 15.19  & 2.17  \\ \hline
128 & 0.9 & 0.001     & 11.68  & 2.2   \\ \hline
128 & 0.9 & 0.0001    & 15.4   & 2.29  \\ \hline
128 & 0.9 & 0.00001   & 15.74  & 2.31  \\ \hline
64  & 0.5 & 0.001     & 16.65  & 3.6   \\ \hline
64  & 0.5 & 0.0001    & 15.42  & 3.79  \\ \hline
64  & 0.5 & 0.00001   & 8.61   & 3.8   \\ \hline
64  & 0.9 & 0.001     & 19.58  & 3.86  \\ \hline
64  & 0.9 & 0.0001    & 8.66   & 3.87  \\ \hline
64  & 0.9 & 0.00001   & 12.76  & 3.94  \\ \hline
Best:& 1.74 & $\mu\pm\sigma:$ & $13.53\pm3.18$  & $2.66\pm0.85$  \\ \hline
\end{tabular}
\caption{Experimental results for TCN model trained with and without P/S phase information on the local subset}
\label{experiments1}
\end{table}

Finally, the TCN model trained on the global STEAD dataset, with and without using P-S phase information, are given in Table \ref{experiments2}. Consistent with our previous finding, the best-performing hyperparameter combinations are shown in bold fonts and have an L1 Loss score of 3.14 km without P/S information as the input, and 2.64 km with its inclusion. The same results are depicted in Figure \ref{TCN_Global}. 

\begin{table}[ht!]
\centering
\begin{tabular}{|c|c|c|c|c|}
\hline
\multicolumn{5}{|c|}{TCN Global} \\
\hline
Size & $\gamma$ & lr & No PS & PS\\
\hline
256 & 0.5 & 0.001     & 10.8  & 3.51  \\ \hline
256 & 0.5 & 0.0001    & 10.18 & 3.48  \\ \hline
256 & 0.5 & 0.00001   & 5.35  & 3.45  \\ \hline
256 & 0.9 & 0.001     & 4.16  & 3.45  \\ \hline
256 & 0.9 & 0.0001    & 3.02  & 3.44  \\ \hline
256 & 0.9 & 0.00001   & 2.88  & 3.43  \\ \hline
128 & 0.5 & 0.001     & 7.73  & 2.96  \\ \hline
128 & 0.5 & 0.0001    & 6.8   & 2.94  \\ \hline
128 & 0.5 & 0.00001   & 5.52  & 2.93  \\ \hline
128 & 0.9 & 0.001     & 5.12  & 2.91  \\ \hline
128 & 0.9 & 0.0001    & 3.21  & 2.9   \\ \hline
128 & 0.9 & 0.00001   & 3.16  & 2.88  \\ \hline
64  & 0.5 & 0.001     & 12.56 & 2.86  \\ \hline
64  & 0.5 & 0.0001    & 9.67  & 2.84  \\ \hline
64  & 0.5 & 0.00001   & 9.32  & 2.81  \\ \hline
64  & 0.9 & 0.001     & 8.55  & 2.74  \\ \hline
64  & 0.9 & 0.0001    & 3.67  & 2.7   \\ \hline
64  & 0.9 & 0.00001   & 3.14  & \textbf{2.64}  \\ \hline
Best:& 2.64 & $\mu\pm\sigma:$ & $6.38\pm3.15$  & $3.05\pm0.31$  \\ \hline
\end{tabular}
\caption{Experimental results for TCN model trained with and without P/S phase information on the global dataset.}
\label{experiments2}
\end{table}

\begin{figure}[t]
    \centering
    \includegraphics[width=0.8\linewidth]{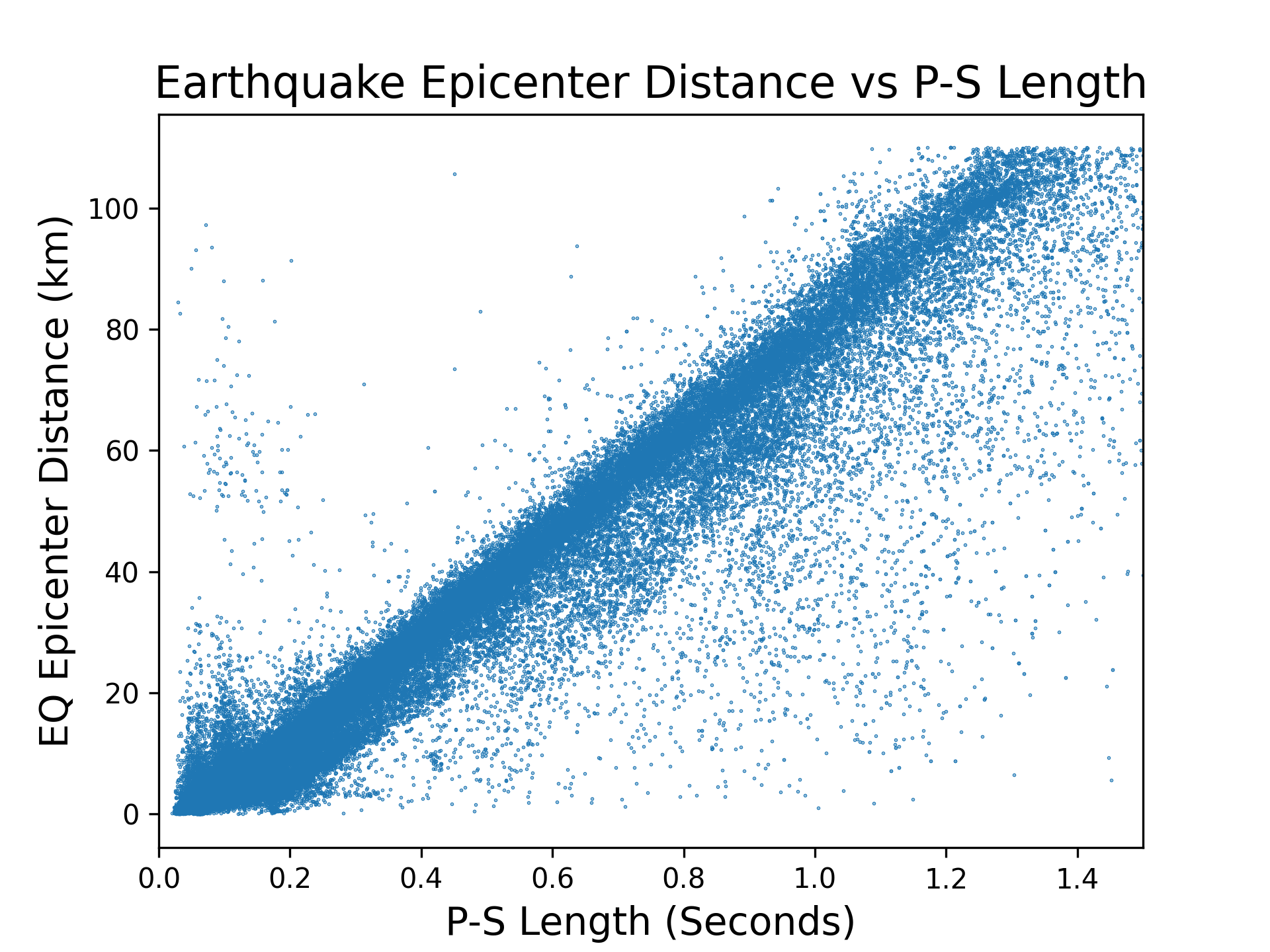}
    \caption{Epicentral Distances (km) vs P/S phase arrival differences (sec.). A clear correlation between P/S phase information and the epicentral distances can be observed.}
    \label{fig:epi_ps}
\end{figure}

\begin{figure*}
  \centering
  \begin{subfigure}{0.49\columnwidth}
    \includegraphics[trim=0 0 0 0,clip=true,width=\textwidth]{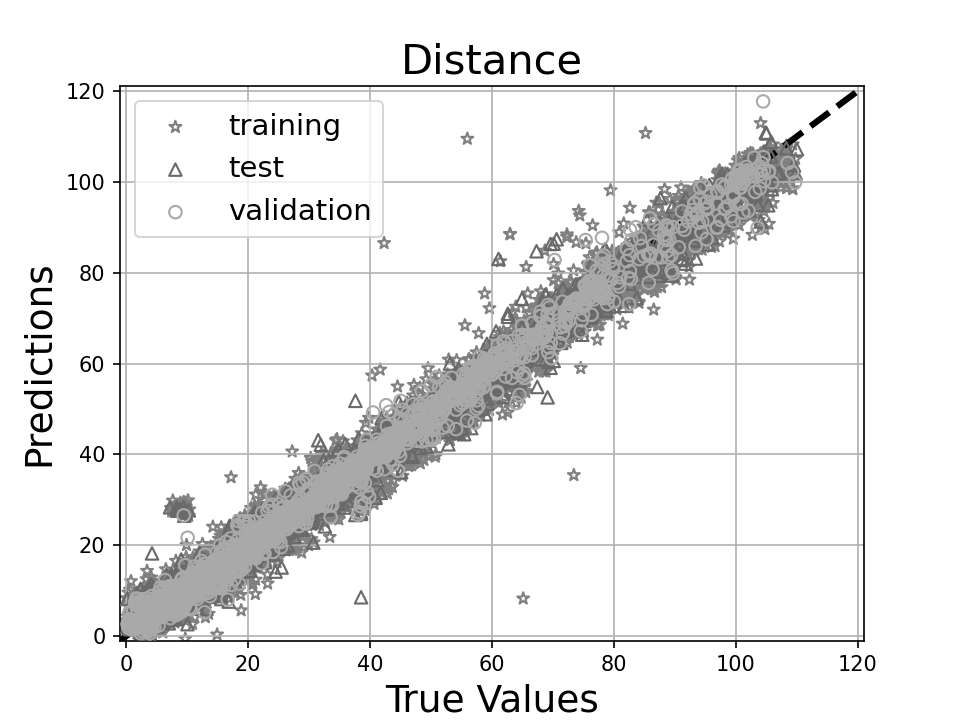}
    \caption{TCN, Local, L1 Loss $2.66\pm0.85$ km}  
    \label{TCN_local_ps}
  \end{subfigure}
  \begin{subfigure}{0.49\columnwidth}
    \includegraphics[trim=0 0 0 0,clip=true,width=\textwidth]{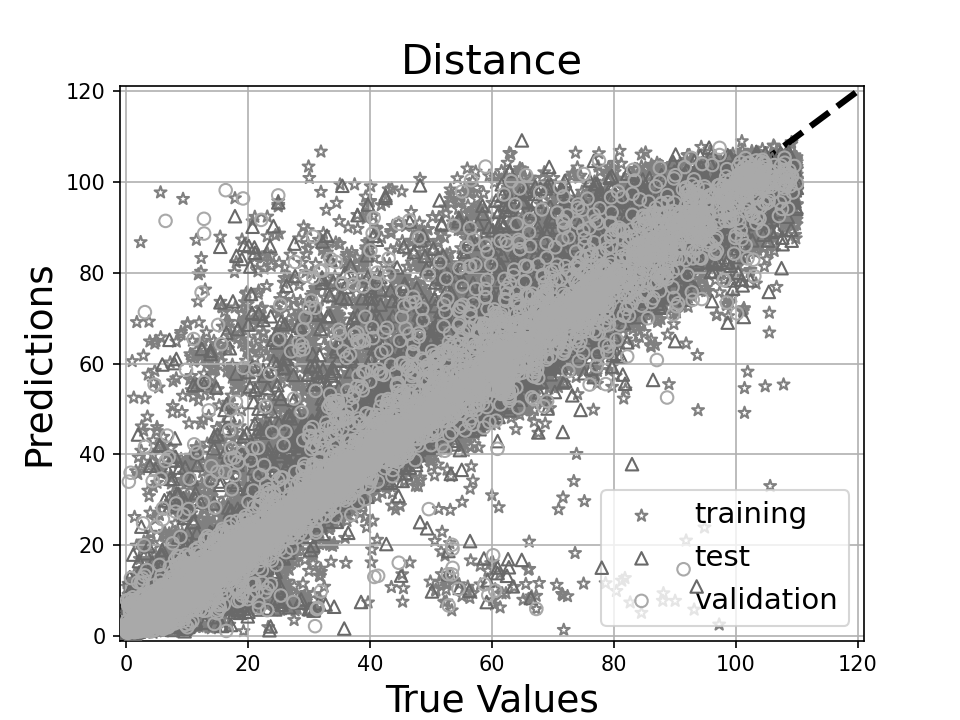}
    \caption{TCN, Global, L1 Loss $3.05\pm0.31$ km}
    \label{TCN_global_ps}
  \end{subfigure}
  \begin{subfigure}{0.49\columnwidth}
    \includegraphics[trim=0 0 0 0,clip=true,width=\textwidth]{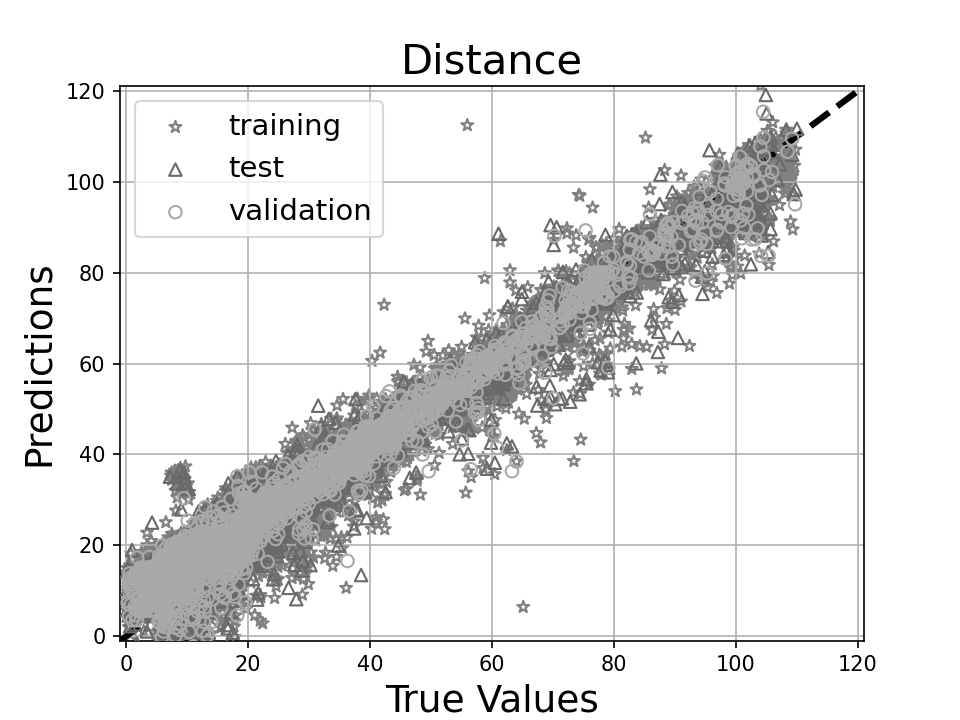}
    \caption{ResNet, Local, L1 Loss $8.28\pm3.69$ km}  
    \label{ResNet_local_ps}
  \end{subfigure}
  \begin{subfigure}{0.49\columnwidth}
    \includegraphics[trim=0 0 0 0,clip=true,width=\textwidth]{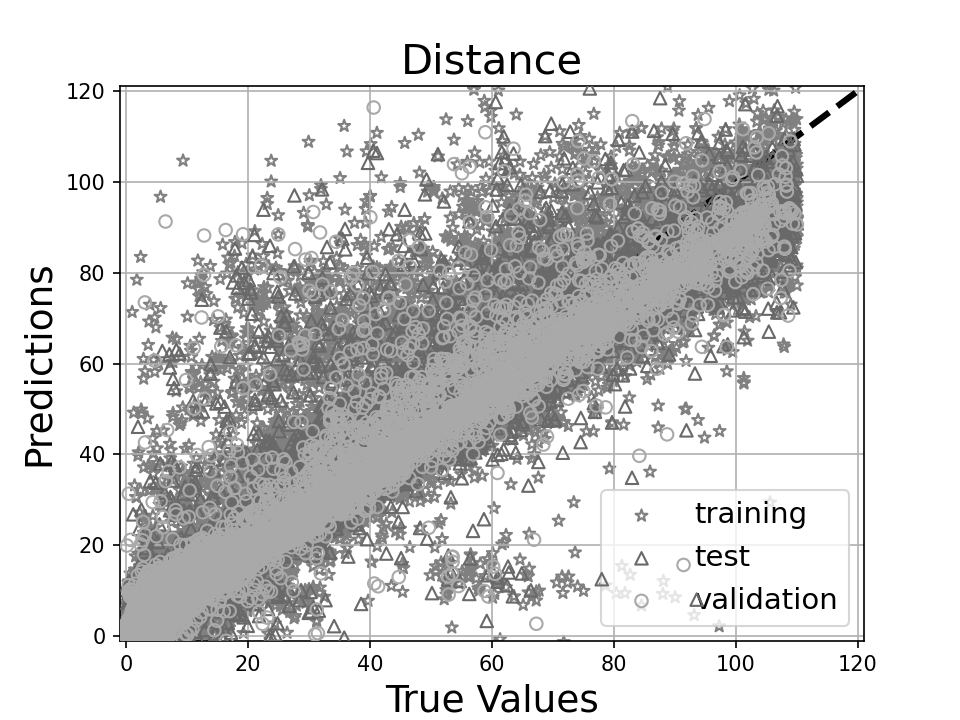}
    \caption{ResNet, Global, L1 Loss $13.02\pm13.11$ km}
    \label{ResNet_global_ps}
  \end{subfigure}
  \caption{Overall summary of the models trained with the P/S phase information in the input. The best model has an L1 Loss score of 2.03km (Train: Star, Test: Triangle, Validation: Circle)}
  \label{distance_sum_ps}
\end{figure*}

\begin{figure*}[t]
  \centering
  \begin{subfigure}{0.49\columnwidth}
    \includegraphics[trim=0 0 0 0,clip=true,width=\textwidth]{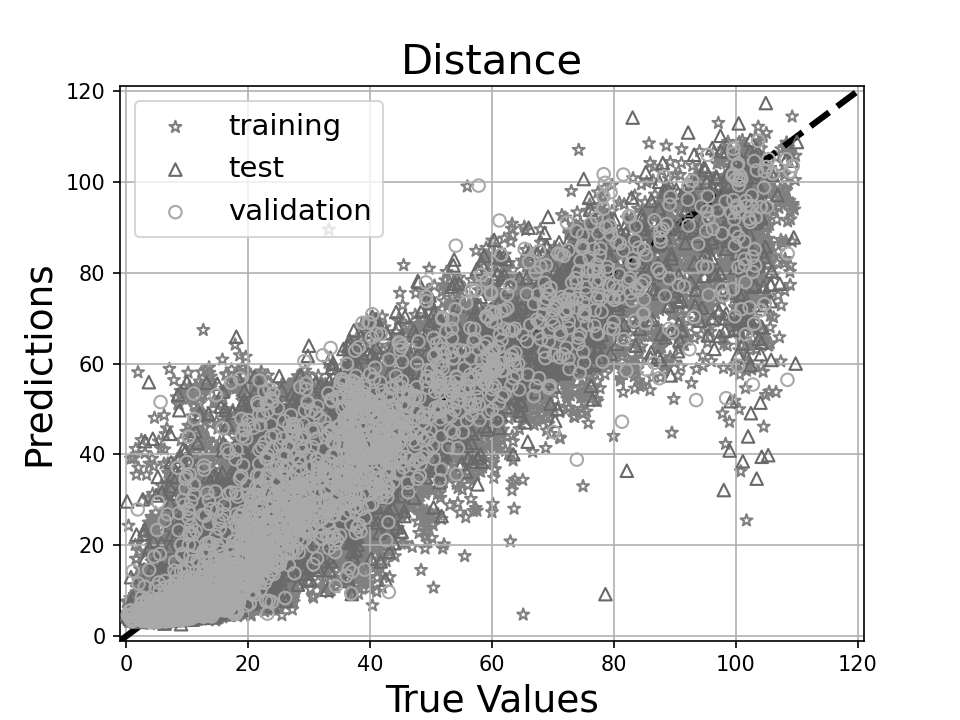}
    \caption{TCN, Local, L1 Loss $13.53\pm3.18$ km}
    \label{TCN_local_nops}
  \end{subfigure}
  \begin{subfigure}{0.49\columnwidth}
    \includegraphics[trim=0 0 0 0,clip=true,width=\textwidth]{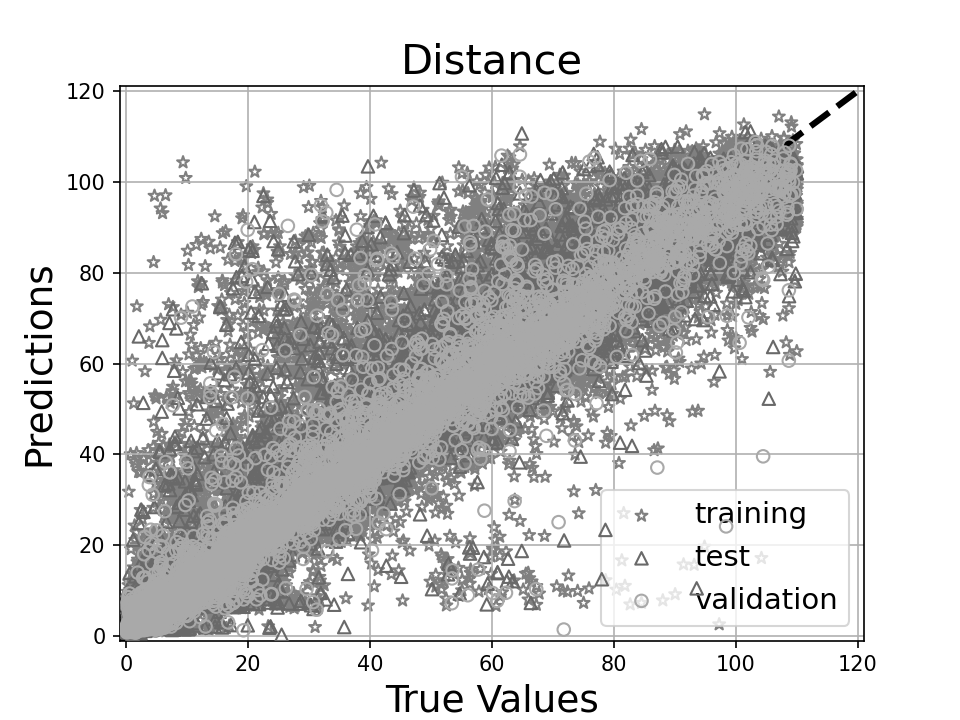}
    \caption{TCN, Global, L1 Loss $6.38\pm3.15$ km}
    \label{TCN_global_nops}
  \end{subfigure}
  \begin{subfigure}{0.49\columnwidth}
    \includegraphics[trim=0 0 0 0,clip=true,width=\textwidth]{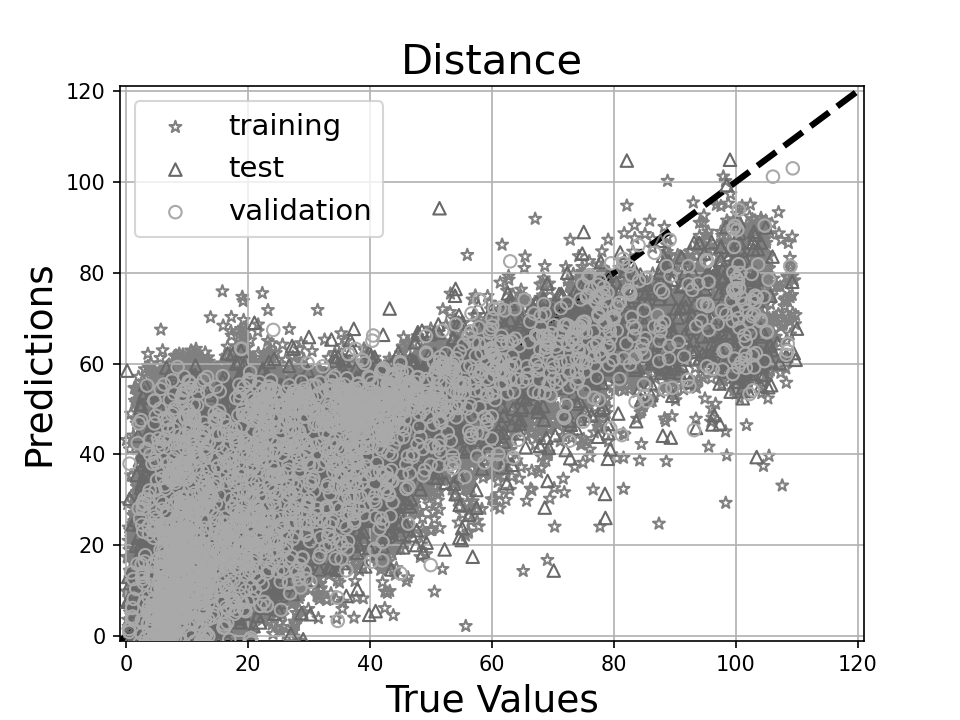}
    \caption{ResNet, Local, L1 Loss $27.67\pm13.74$ km}
    \label{ResNet_local_nops}
  \end{subfigure}
  \begin{subfigure}{0.49\columnwidth}
    \includegraphics[trim=0 0 0 0,clip=true,width=\textwidth]{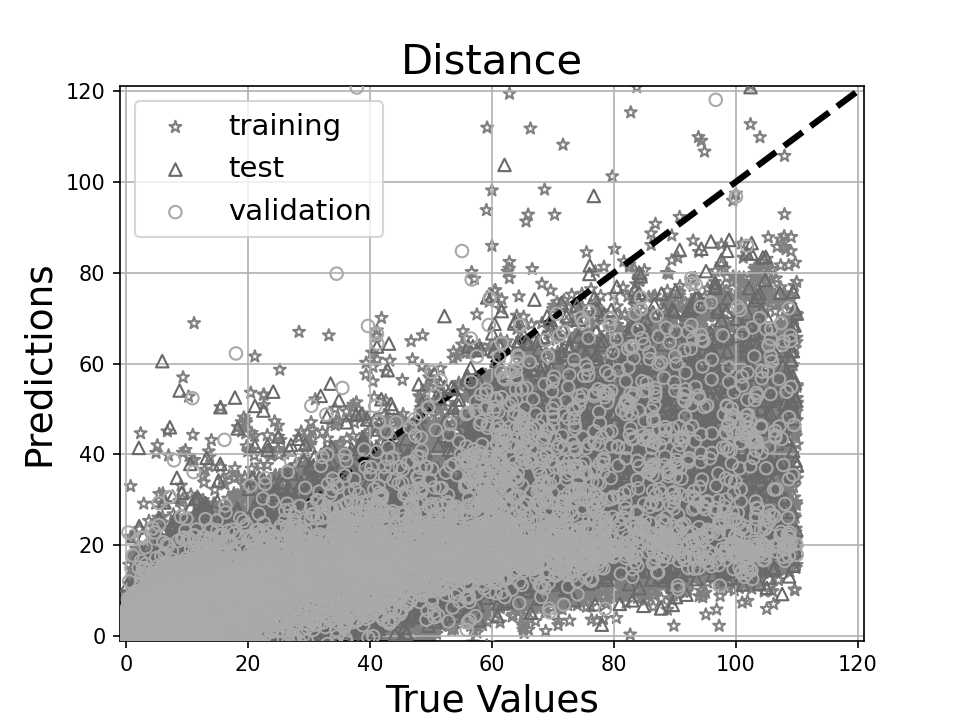}
    \caption{ResNet, Global, L1 Loss $21.58\pm4.02$ km}
    \label{ResNet_global_nops}
  \end{subfigure}
  \caption{Overall summary of the models trained without the P/S phase information in the input. The best model has an L1 Loss score of 3.02km (Train: Star, Test: Triangle, Validation: Circle).}
  \label{distance_sum_nops}
\end{figure*}
The experiments in this section again agree with our hypothesis that P/S phase information has a distinct effect on the performance of deep learning models for epicenter distance prediction from single station recordings. For the TCN models, we have achieved the best L1 Loss scores of 1.74 km and 2.64 km, for local and global sets, respectively. We observe that the inclusion of the P/S phase information significantly drops the loss score by up to four times when compared to 7 km (local set) and 3.14 km (global set) L1 Loss scores obtained without the P/S phase information.

\subsection{Experimentation Summary}

Overall, we used two deep learning models based on two fundamental convolutional architectures, namely RNN and TCN, both global and local subsets of STEAD. P/S phase information was included as an additional (i.e. 4th) input channel for ablation experiments.
The results demonstrate a strong correlation between P/S phase information and the epicentral distances. Based on the model and dataset, the training time increases by about 6\%, and the losses improve between 114\% and 343\% when P/S information is included as an input channel. A summary of all results are provided in Table \ref{experiment_summary}. In Figure \ref{TCN_global_nops}, where the best-performing model obtained without the incorporation of P/S phase information is depicted, we observe a partial diagonal scatter. Because this experiment is carried out on the global set with the better performing model, TCN, this underscores the significance of having access to large-scale data for the effectiveness of deep learning methodologies.

\begin{table}[ht!]
\centering

\begin{tabular}{|l|l|l|l|l|l|l|}
\hline
Signal & Model  & Data    & Train      & Val      & Test    & Runtime \\ \hline
P-S    & TCN    & Local   & 1.85       & 1.65     & 1.74    & 85      \\ \hline
P-S    & TCN    & Global  & 2.83       & 2.78     & 2.64    & 344     \\ \hline
P-S    & ResNet & Local   & 1.42	   & 2.16	  & 4.47	& 31      \\ \hline
P-S    & ResNet & Global  & 2.54	   & 4.03	  & 4.31	& 128     \\ \hline
No P-S & TCN    & Local   & 5.88	   & 5.45	  & 7.00	& 82      \\ \hline
No P-S & TCN    & Global  & 3.13	   & 3.05	  & 3.02	& 338     \\ \hline
No P-S & ResNet & Local   & 5.02	   & 10.97	  & 13.33	& 29      \\ \hline
No P-S & ResNet & Global  & 5.10	   & 12.96	  & 14.82	& 118     \\ \hline
\end{tabular}
\caption{Summary of the experiments with best results in each category. Train and Validation and Test performance metrics are in L1 Loss, the runtime is in minutes.}
\label{experiment_summary}
\end{table}

\subsection{Correlation between the P/S Phase Information and the Epicentral Distance}

As stated earlier, the epicentral distance is typically estimated by measuring the time difference between the arrival of P waves and S waves at a station. The greater the time difference between the waves, the farther the epicenter is from that station. 

The effects of the inclusion the P/S phase information on the performance metrics for different models, subsets and various hyperparameters can be observed in Table \ref{experiment_summary}. We utilize both Pearson's linear correlation and Spearman's rank correlation techniques to compute the correlation between epicentral distances and P-to-S phase intervals, yielding coefficients of 0.956 and 0.926, respectively. The epicentral distance and P-to-S phase intervals for each individual event of the global set are visualized in Figure \ref{fig:epi_ps}, which indicates a linear relationship.

\begin{figure*}[t]
  \centering
  \begin{subfigure}{0.49\columnwidth}
    \includegraphics[trim=0 0 0 0,clip=true,width=\textwidth]{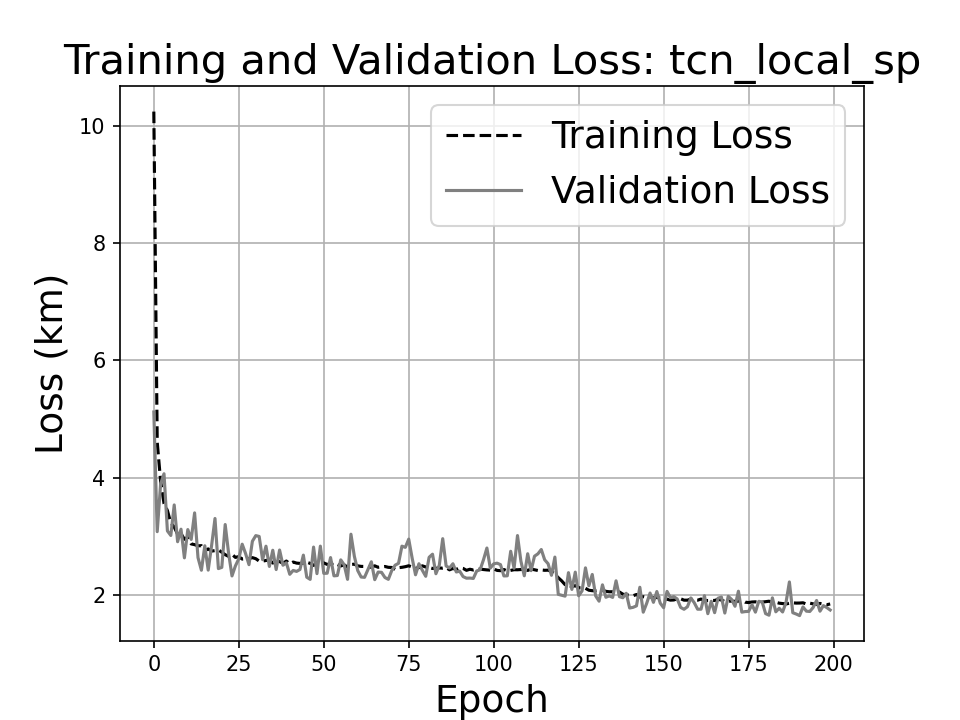}
    \caption{TCN, Local, L1 Loss 1.74}
  \end{subfigure}
  \begin{subfigure}{0.49\columnwidth}
    \includegraphics[trim=0 0 0 0,clip=true,width=\textwidth]{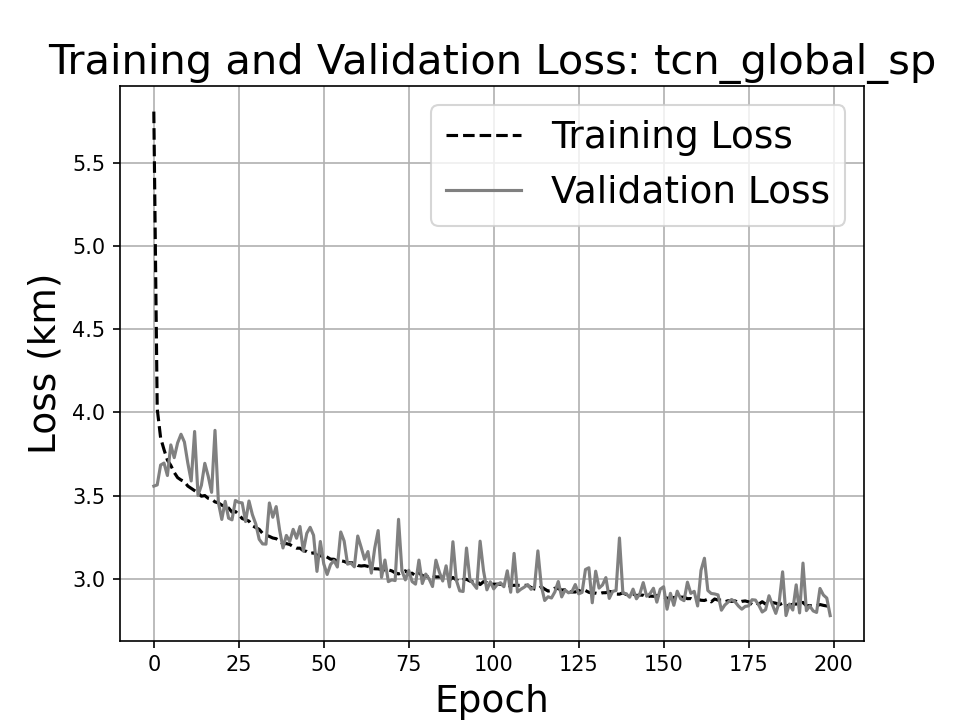}
    \caption{TCN, Global, L1 Loss 2.64}
  \end{subfigure}
  \begin{subfigure}{0.49\columnwidth}
    \includegraphics[trim=0 0 0 0,clip=true,width=\textwidth]{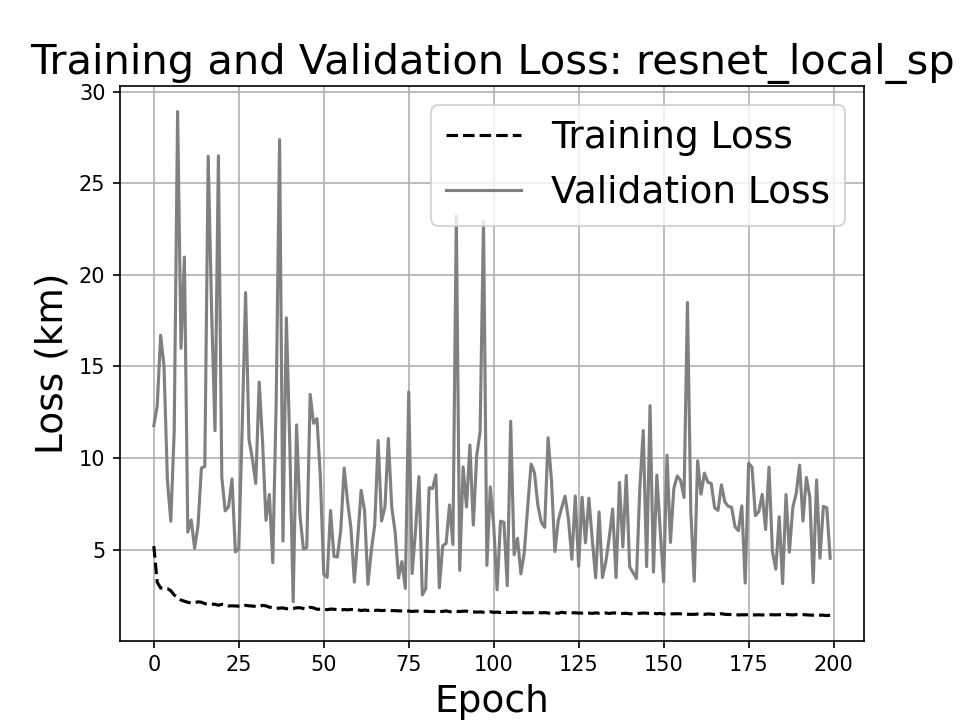}
    \caption{ResNet, Local, L1 Loss 4.47}
  \end{subfigure}
  \begin{subfigure}{0.49\columnwidth}
    \includegraphics[trim=0 0 0 0,clip=true,width=\textwidth]{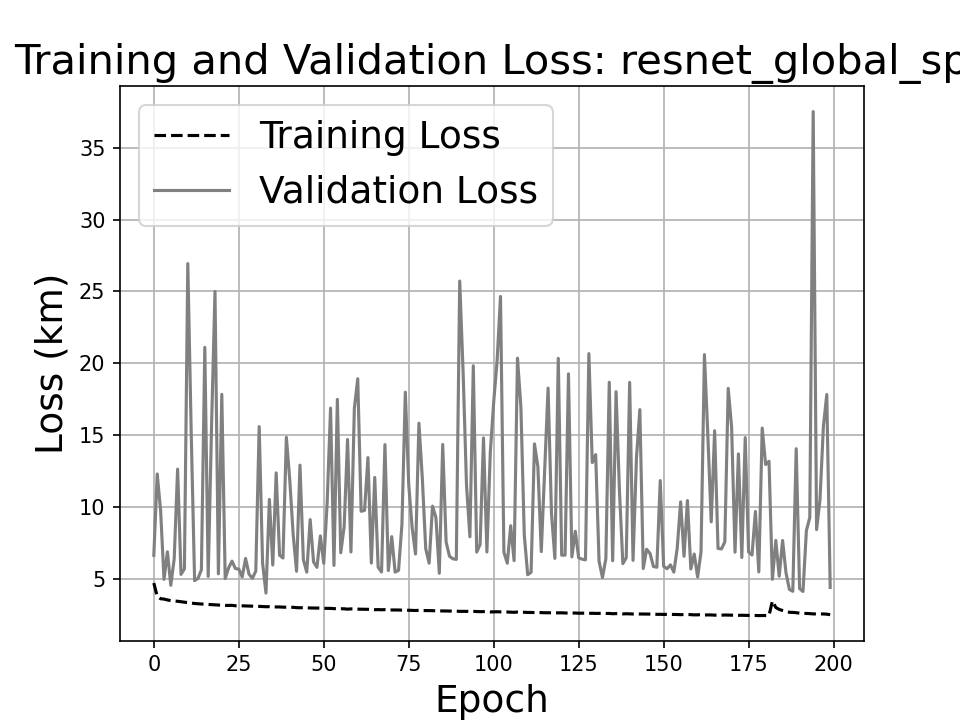}
    \caption{ResNet, Global, L1 Loss 4.31}
  \end{subfigure}
  \caption{Overall learning curves of the models trained with P/S phase information. Best model Loss score: 1.74.}
  \label{lc_sum_ps}
\end{figure*}
\begin{figure*}[t]
  \centering
  \begin{subfigure}{0.49\columnwidth}
    \includegraphics[trim=0 0 0 0,clip=true,width=\textwidth]{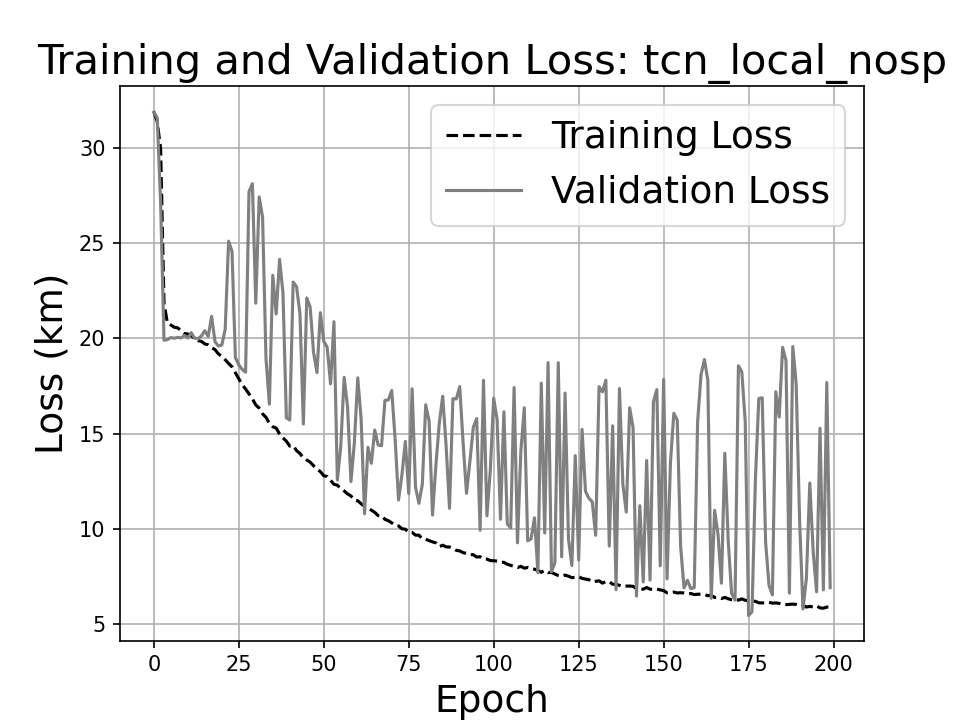}
    \caption{TCN, Local, L1 Loss 6.91}
  \end{subfigure}
  \begin{subfigure}{0.49\columnwidth}
    \includegraphics[trim=0 0 0 0,clip=true,width=\textwidth]{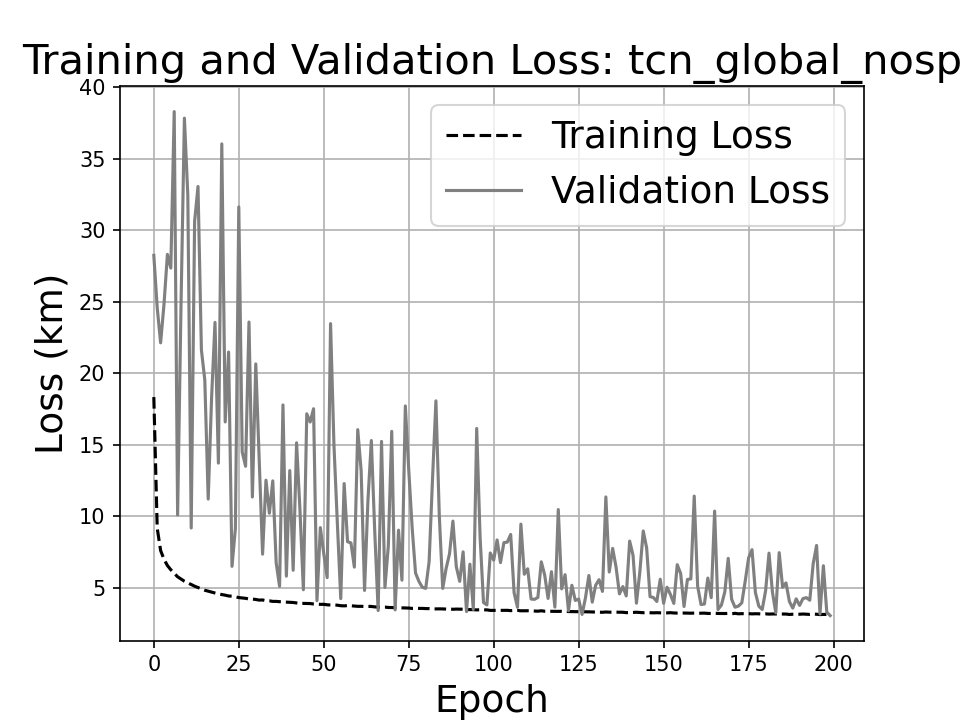}
    \caption{TCN, Global, L1 Loss 3.02}
  \end{subfigure}
  \begin{subfigure}{0.49\columnwidth}
    \includegraphics[trim=0 0 0 0,clip=true,width=\textwidth]{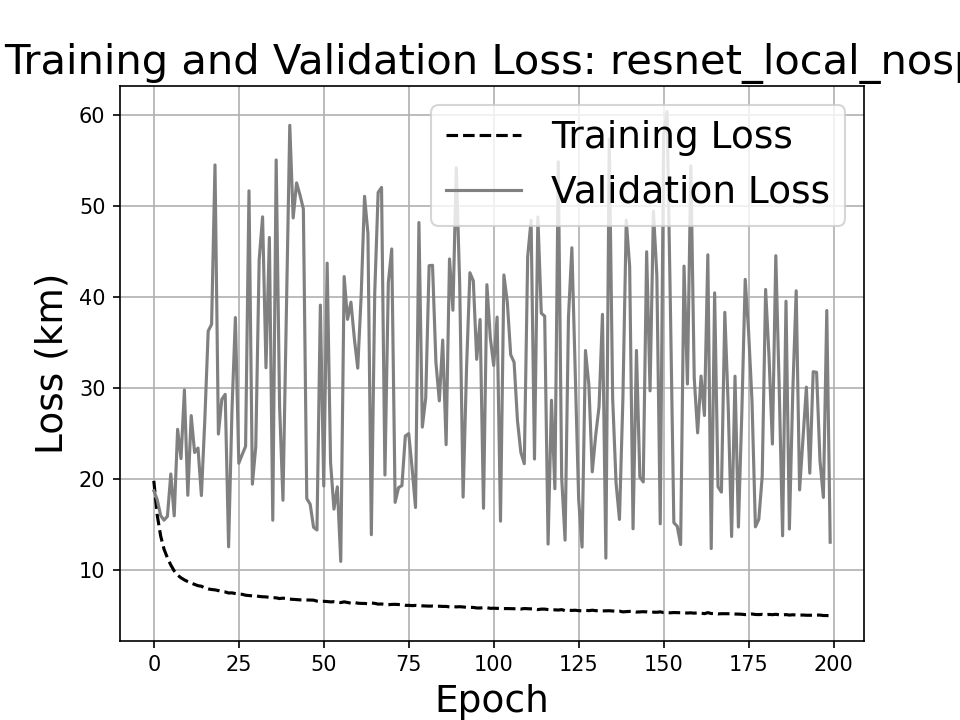}
    \caption{ResNet, Local, L1 Loss 13.33}
  \end{subfigure}
  \begin{subfigure}{0.49\columnwidth}
    \includegraphics[trim=0 0 0 0,clip=true,width=\textwidth]{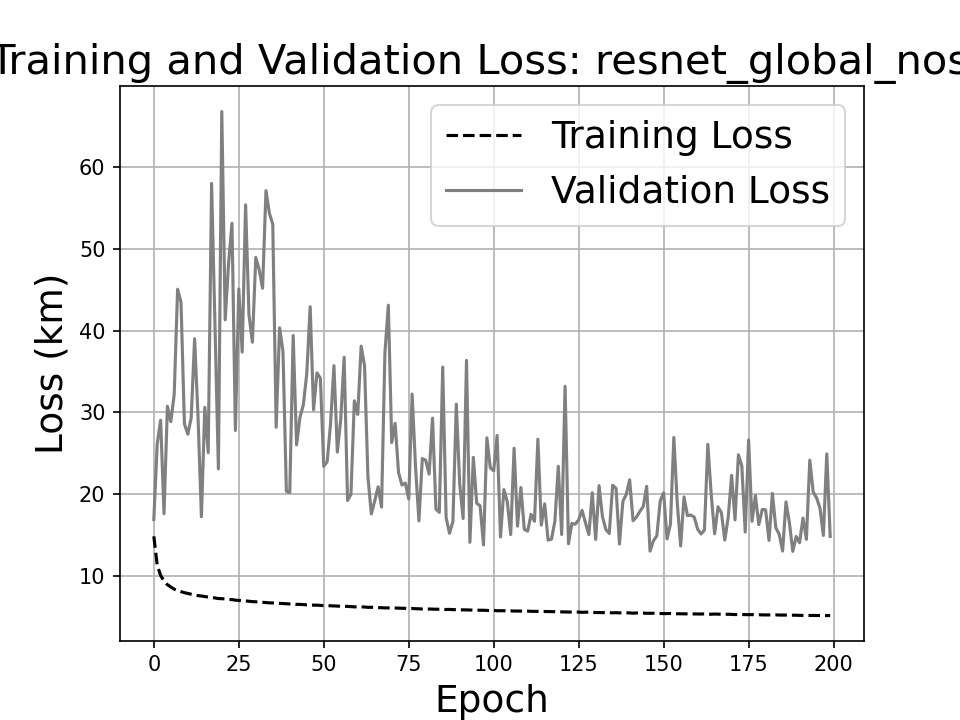}
    \caption{ResNet, Global, L1 Loss 14.82}
  \end{subfigure}
  \caption{Overall learning curves of the models trained without the P/S phase information. Best model Loss score: 3.02 km.}
  \label{lc_sum_nops}
\end{figure*}

\subsection{Distance Prediction Capability and Training Plots}

The effects of different hyperparameters, especially the inclusion of P/S phase information, is visualized in Figure \ref{distance_sum_ps}, for the best-performing models in each category. The prediction vs. ground truth graphs show the training (red), validation (green) and test (test) results for each event, with mostly a diagonal-like scatter. Compared to the experiments without P/S information inputs, as shown in Figure \ref{distance_sum_nops}, the predictions are observable better for a majority of events.  Learning curves for the corresponding experiments are given for the experiments with and without the P/S phase information, in Figures \ref{lc_sum_ps} and \ref{lc_sum_nops}, respectively. The training plots for deep learning models highlight an important difference between the presence and absence of P-S signal for the aforementioned models. The models trained without P-S signal tend to have higher discrepancy of the epicentral distance estimation performance for validation and training sets. We conclude that the presence of P-S signal prevent the overfitting of deep learning models and enable more robust feature extraction, resulting in better generalizability.

\clearpage
\clearpage

\section{Conclusion}
In this study, we conduct a hyperparameter search on two deep learning models, TCN and ResNet, to explore their efficacy in learning from ground motion records, while also investigating the influence of auxiliary information on model performance. We found that by incorporating arrival times of Primary (P) and Secondary (S) earthquake waves, known as the P/S phases, models demonstrated improved performance. Furthermore, our analyses reveal a strong correlation between the epicentral distances and P/S phase information, supported by Pearson's and Spearman's correlation coefficients of 95.6\% and 92.6\%, respectively. Our findings indicate a challenge in deep learning from accelerometer signals. The models tend to rely heavily on the highly correlated P/S phase information. ResNet models tend to have higher oscillations in the validation scores, TCN models usually have validation scores on par with training scores and with much less oscillations. This oscillating behaviour in training plots, is also evident with the increased standard deviation for models trained without P/S phase information or for ResNet models in general. Thanks to our extensive model training approach, we can conclude that TCN models are better suited for this task than ResNet models, and inclusion of P/S phase information improves training capability of models, decreasing the overfitting behaviour of deep learning models, and improves the overall model performance, achieving a better performance and generalizability.

To the best of our knowledge, there is a notable absence of deep learning studies exploring the impact of auxiliary information on model performance in seismic data analysis. In our study, we investigated the influence of P/S phase information. However, auxiliary information could extend to factors such as the distribution of seismic stations within a network when multiple station locations are incorporated into the network, which is why we work on records from single stations. While deep learning in seismology and earthquake engineering methodologies show promising results, our study suggests that models may struggle to capture the nuanced characteristics of ground motion recordings.

Our study underscores the need for future research to explore alternative architectures and experimental designs tailored to localized seismic data. The disparity in performance between global and localized subsets suggests avenues for further investigation, potentially leveraging datasets such as local and dense ground motion recordings. By refining our methodologies, we aim to advance our understanding of seismic phenomena and enhance the efficacy of earthquake epicentral distance estimation techniques. Ultimately, our goal is to develop models capable of extracting high-level features from ground motion records, thus contributing to the advancement of AI-based seismic data analysis.

\section*{Acknowledgments}
This work is supported by The Scientific and Technological Research Council of Turkey (TÜBİTAK) as a part of the ongoing TÜBİTAK 1001 funded project, Project No.121M732, titled "Deep Learning and Machine Learning Based Dynamic Soil and Earthquake Parameter Estimation Using Strong Ground Motion Station Records”. 

The numerical calculations reported in this paper were fully performed at TUBITAK ULAKBIM, High Performance and Grid Computing Center (TRUBA).

The authors would like to thank Dr. S. Mostafa Mousavi, who is the first author of \cite{Mousavi2020a} and a contributor to the STEAD, for sharing the details of their experimental setup, so that an accurate benchmark could be conducted.

\section*{Computer Code Availability}

The computer code used in this work is available at github.com/caglarmert/mage. Corresponding Author can be contacted for inquiries. The experiments reported in this work is also available online at api.wandb.ai/links/caglarmert/rdvjvsyu. Experiments were done with NVIDIA GTX 3070 and A100 GPUs.

\bibliographystyle{plain}
\bibliography{bibfile}

\end{document}